# Imaging of anomalous internal reflections of hyperbolic phonon-polaritons in hexagonal boron nitride

Alexander J. Giles,[†] Siyuan Dai,[‡] Orest J. Glembocki,[#] Andrey V. Kretinin,[¶] Zhiyuan Sun,[‡] Chase T. Ellis,[†] Joseph G. Tischler,[#] Takashi Taniguchi,[§] Kenji Watanabe,[§] Michael M. Fogler,[‡] Kostya S. Novoselov,[¶] Dimitri. N. Basov,[‡,∥] and Joshua D. Caldwell[*,#]

[†]*NRC Postdoctoral Fellow residing at the U.S. Naval Research Laboratory, 4555 Overlook Ave, S.W., Washington, D.C., USA*
[#]*U.S. Naval Research Laboratory, 4555 Overlook Ave, S.W., Washington, D.C., USA*
[‡]*Department of Physics, University of California San Diego, 9500 Gilman Drive, La Jolla, California 92093, USA*
[¶]*School of Physics and Astronomy, University of Manchester, Oxford Rd, Manchester, UK*
[§]*National Institute for Materials Science, 1-1 Namiki Tsukuba, Ibaraki, Japan*
[∥]*Department of Physics, Columbia University, 538 West 120th Street, New York, New York 10027*

E-mail: joshua.caldwell@nrl.navy.mil

## Abstract

We use scanning near-field optical microscopy to study the response of hexagonal boron nitride nanocones at infrared frequencies, where this material behaves as a hyperbolic medium. The obtained images are dominated by a series of "hot" rings that occur on the sloped sidewalls of the nanocones. The ring positions depend on the incident laser frequency and the nanocone shape. Both dependences are consistent with directional propagation of hyperbolic phonon-polariton rays that are launched at the edges and zigzag through the interior of the nanocones, sustaining multiple internal reflections off the sidewalls. Additionally, we observe a strong overall enhancement of the near-field signal at discrete resonance frequencies. These resonances attest to low dielectric losses that permit coherent standing waves of the sub-diffractional polaritons to form. We comment on potential applications of such shape-dependent resonances and the field concentration at the hot rings.

**Introduction.** Hyperbolic media (HM)[1] have attracted much attention for their unusual optical properties that enable confinement of electromagnetic energy to extremely small volumes[2,3] and super-resolution focusing and imaging.[4-13] The discovery[3,14] that hexagonal boron nitride (hBN) is a natural, low-loss HM has led to a surge in interest in this compound[15-17] and its other allotropes.[18,19] The term hyperbolic refers to materials or metamaterials where the dielectric functions along orthogonal crystal axes are opposite in sign. In the case of hBN, this is realized in two distinct spectral Reststrahlen bands as shown in Figure 1a, with both type I ($\varepsilon_t > 0$, $\varepsilon_z < 0$) hyperbolicity in the lower band, $\omega = 760\text{–}825 \text{ cm}^{-1}$ (henceforth, the L-band) and type II ($\varepsilon_t < 0$, $\varepsilon_z > 0$) behavior in the upper band, $\omega = 1360\text{–}1610 \text{ cm}^{-1}$ (the U-band). The negative signs of the permittivity tensor components $\varepsilon_t$, $\varepsilon_z$ arise due to the strong phonon resonances[20] of this highly anisotropic polar material, with the subscripts $t$ and $z$ referring to the transverse ($x,y$) and axial ($z$) directions, respectively. Low optical losses (compared to those reported for artificial hyperbolic metamaterial structures) make hBN well suited for fundamental investigations of its collective modes, hyperbolic phonon-polaritons (HPhPs), which are the extraordinary rays of this uniaxial material. The isofrequency surfaces of these modes, defined by the equation[1, 9] $k_t^2/\varepsilon_z + k_z^2/\varepsilon_t = (\omega/c)^2$ are open (Figure 1, b and c), implying that their momentum $\mathbf{k} = (\mathbf{k}_t, k_z)$ can nominally be arbitrarily large (the upper limit is the size of the Brillouin zone of hBN). In particular, $k = |\mathbf{k}|$ can greatly exceed the photon momentum $\omega/c$ in vacuum, in which case the isofrequency surfaces can be approximated by the cones:

$$\varepsilon_t(\omega)k_t^2 + \varepsilon_z(\omega)k_z^2 = 0. \tag{1}$$

The group velocity $\mathbf{v} = \partial_\mathbf{k}\omega$ of such polaritons is nearly orthogonal to the phase velocity (Figure 1, b and c) and thus its propagation is directionally restricted. Namely, the HPhPs will propagate at a fixed angle, either $\theta$ or $180° - \theta$, with respect to the optical axis, where

$$\theta(\omega) = \arctan\frac{k_z}{k_t} = 90° - \arctan\frac{\sqrt{\varepsilon_t(\omega)}}{i\sqrt{\varepsilon_z(\omega)}}. \tag{2}$$

From the $\omega$-dependence of $\varepsilon_t$, $\varepsilon_t$ (plotted in Figure 1a) one concludes that $\theta(\omega)$ has the opposite behavior in the two hyperbolic bands of hBN.[3-5] It increases with $\omega$, from 90° to 180°, across the L-band, but decreases from 90° to 0°, across the U-band. These unusual and unique properties of phonon-polaritons in hBN strongly contrast with those of more familiar surface-bound polaritons in typical metals,[21] semiconductors,[22] and polar dielectrics.[23-25]

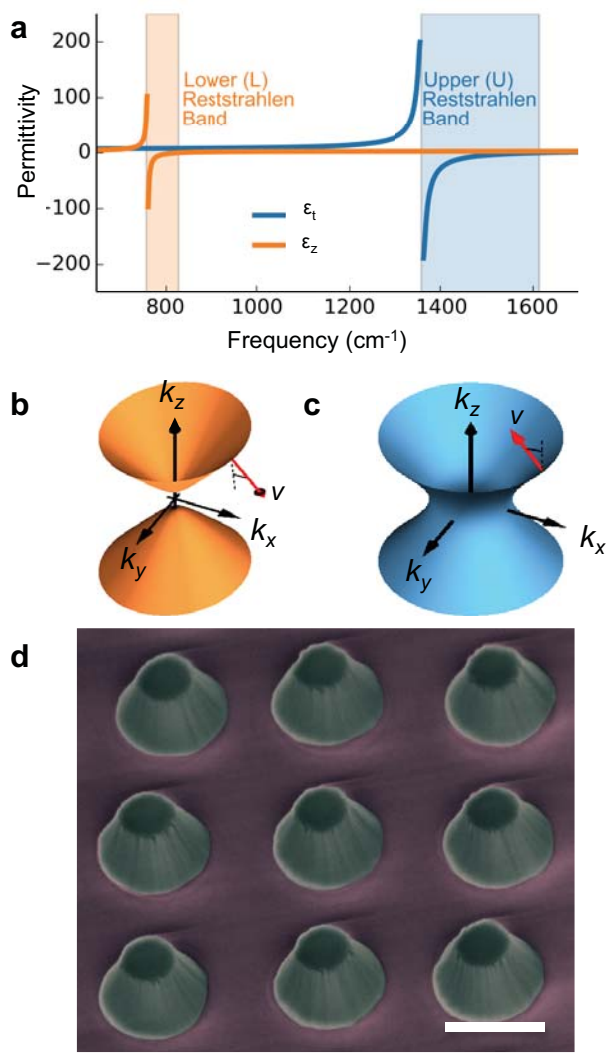

**Figure 1. a** Derived real parts of the permittivity tensor components of hexagonal boron nitride. Schematic isofrequency contours for a **b** type I and **c** type II hyperbolic material. The arrows indicate the directions of the group velocity (*v*) of the hyperbolic polariton modes with respect to the optic axis. **d** Tilted scanning electron microscope image of a representative array of hBN nanocones on a silicon substrate. Scale bar is 500 nm.

For a qualitative understanding of our experimental results, it is important that the confinement and propagation of HPhPs inside three-dimensionally confined hBN nanocones be visualized. In particular, it is imperative that the total internal reflections these volume-confined HPhP modes experience (at the nanostructure surfaces) be understood. Such a reflection is anomalous if the surface is tilted with respect to the optical (*z*) axis, as is the case for the nanocones explored here (Figure 2). Under such conditions, the incident and reflected angles are not equal, and the corresponding momenta $k_1$, $k_2$ and group velocities $v_1$, $v_2$ are nonequivalent as well (Figure 2a).

In addition to this unusual ray optics, one also expects polaritons to exhibit novel wave-like properties, e.g., long-lived, discrete eigenmodes (resonances) in hBN nanoparticles, in the presence of low enough dielectric losses. Experimental evidence for such resonances has been obtained in a previous work[3] by measuring far-field reflectivity of hBN nanocone arrays. In this Letter we report on a near-field imaging study of individual nanocones from a subset of those samples. We present detailed maps of the electric field distribution on and off resonance that provide direct evidence that the aforementioned anomalous internal reflections (Figure 2b) of the HPhPs continue, even after multiple internal reflections. This is in contrast to the specular reflection anticipated for non-hyperbolic media (Figure 2c).

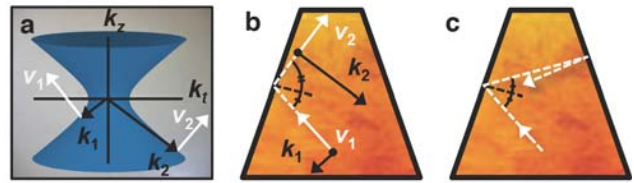

**Figure 2.** Anomalous internal reflection of polariton rays. **a** Frequency-momentum diagram for a type II hyperbolic material. Group velocities are depicted with white arrows while the momentum vectors are depicted with black arrows. **b** Real-space schematic showing anomalous reflection inside a nanocone cross section. The propagation angle is defined by the vectors shown in Figure 2a and is independent of the sidewall slope. **c** Specular reflection, where the propagation direction is dictated by the sidewall slope.

The two samples studied in our experiments contained periodic arrays of hBN truncated nanocones fabricated on a high resistivity, nominally undoped Si substrate. First, relatively large flakes of hBN were deposited onto the substrate by mechanical exfoliation. Then the array of nanocones was formed by etching in a $CHF_3/O_2$ inductively coupled plasma (ICP) using electron beam patterned aluminum circles as an etching mask. The diameter of the circles defined the truncated (top) radius of the cones and the conical shape originated from the self-passivation in the $CHF_3$-based plasma. After reactive ion etching, the aluminum mask was stripped in a standard wet aluminum etchant.[3] A tilted SEM image of the nanocone array making up Sample 1 is presented in Figure 1d, with each of the cones having a height of 360 nm, a median diameter of 330 nm (corresponding to an aspect ratio of 0.92), and a sidewall angle of 27°. The center-to-center distance of these nanocones was 600 nm and the total dimensions of the array was $50 \times 50\ \mu m^2$. Sample 2 consisted of a 330 nm tall, 365 nm median diameter nanocones (aspect ratio 1.17), a sidewall angle of 25° and a center-to-center distance of 1000 nm. Far-field reflection spectra from various arrays of the same size, but with varying periodicity indicated that near-field coupling between adjacent pillars was negligible for both samples. While these and the other samples explored in previous work[3] were originally designed and fabricated for far-field optical measurements, they are also well-suited for our present goal, which is to explore the near-field response of individual nanocones. The angled shape of the nanocones enabled investigation of the optical near-fields both at the top of the nanocones and also along their sidewalls. In addition, having both far- and near-field data for each array proved beneficial as it revealed an intriguing shift between the near- and far-field spectra we discuss below.

Our experimental study was conducted using scattering-type scanning near-field optical microscopy (s-SNOM). Our commercial s-SNOM apparatus (Neaspec) utilized a sharp metallized tip of an atomic force microscope (AFM) as an antenna that scattered incident infrared (IR) light onto the sample. The $p$-polarized infrared laser light was focused onto the tip-sample region by an off-axis parabolic mirror, which was also used to collect the backscattered light. The AFM was operated in tapping mode with an oscillation frequency of 280 kHz and amplitude 70 nm. The pseudo-heterodyne module of the s-SNOM performed interferometric detection of the IR light scattered by the tip. Demodulation of the signal at the third harmonic of the tapping frequency yielded us the amplitude $s_3$ and phase $\phi_3$ data, which represented the genuine near-field electromagnetic response of the sample. The IR sources employed in our experiments included tunable monochromatic quantum cascade lasers from Daylight Solutions (1310–1440 cm$^{-1}$ and 1485–1740 cm$^{-1}$) and a broadband difference-frequency generation laser with a tunable central frequency of 700–2200 cm$^{-1}$ and bandwidth of 400 cm$^{-1}$ (Lasnix). With the latter of these sources we carried out nano-FTIR (Fourier transform IR) spectroscopy in both hyperbolic bands of hBN. Using the former two laser sources, we carried out spatial imaging ($s_3$ and $\phi_3$) on the nanocone surface using the monochromatic s-SNOM. Due to the lack of a single frequency, line-tunable laser source in the L-band, spatial imaging was not possible. However, we were able to perform such nanoimaging over a significant portion of the U-band, which constitute our most striking data that we discuss next.

As the base of the hBN nanoparticle was attached to the substrate, only the top and sidewalls of the nanocones were accessible to our s-SNOM probe. However, we were able to acquire the near-field signal over almost this entire exposed surface. A subset of the obtained images is shown in Figure 3. The three rows in this figure, labeled $^{U}TM_{13}$, $^{U}TM_{14}$, and $^{U}TM_{15}$, correspond to three different resonance frequencies $\omega$ of the nanocone that occur inside the U-band, with the nomenclature $^{R}TM_{ml}$ referring to these transverse magnetic modes in either the R = U- or L-bands, with $z$-axis angular momentum and orbital index, $m$ and $l$, respectively, as described in previous work.[3] (These resonances are observed as peaks in both the far- and near-field spectra, see below.) Each of the rows in Figure 3 displays the maps of $s_3$ (left) and $\phi_3$ (right) at a particular $\omega$, as labeled. The pixel positions in these images are projected onto the $x$–$y$ plane, (the pixel size is 5 nm).

From the spatial maps presented in Figure 3, it is observed that at all frequencies within the U-band that the top surface of the nanocone is the darkest portion of the $s_3$ spatial plots, implying that the signal there is the lowest amplitude. In contrast, the signal amplitude on the sidewalls is much higher. Remarkably, the regions of the strongest intensity form a series of rings about the sidewalls. This is the most conspicuous and novel aspect of our data. Within each successive resonance shown in Figure 3, we find that the number of "hot" rings increases by one, and when compared to the orbital index for any given mode, it is found that the number of rings is equal to $l$. The rings occur not only in $s_3$ but also in $\phi_3$, (see the right column of Figure 3). To accentuate the positions of the rings we have added numerical labels and dashed arcs to each of the images. The difference in diameters of the rings implies that they are situated at different heights from the base of the nanocone. We refer to these high-intensity regions as 'rings' even though they are distorted or interrupted on the lower left and right parts of each image. In fact, these distortions, which are especially prominent near the base of the nanocone, are artifacts caused by the edges of the slightly lopsided, pyramidal AFM tip. Rotating the sample and changing the tip scanning direction in subsequent measurements did not cause a reorientation of the distorted features, verifying this hypothesis.

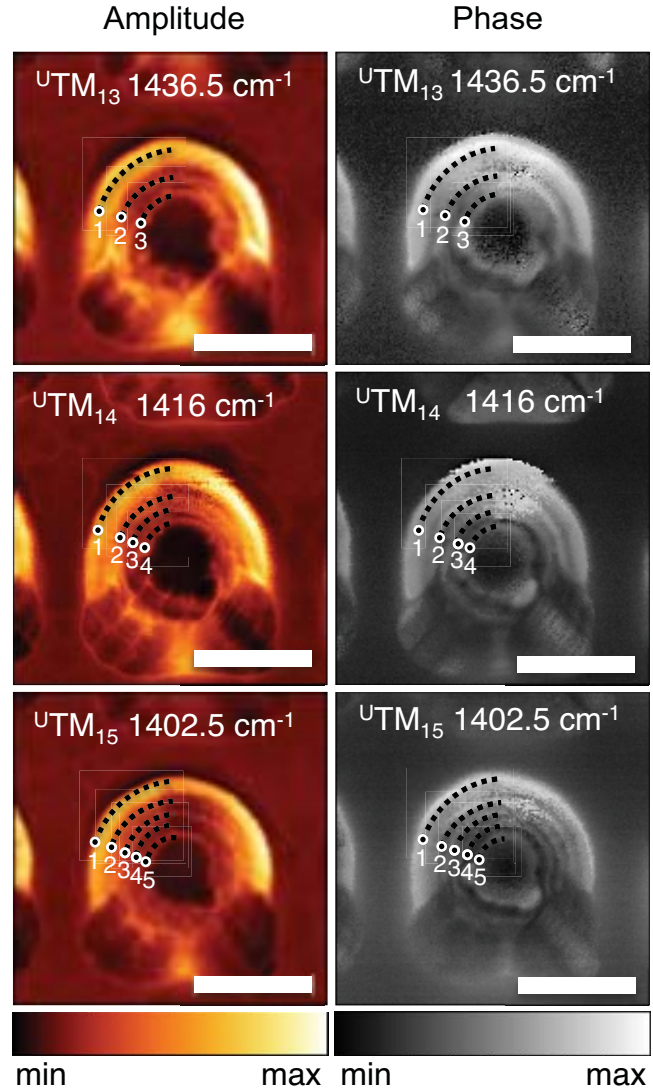

**Figure 3.** On-resonance s-SNOM images obtained for an hBN nanocone from Sample 1. The false-color and grayscale maps show, respectively, the amplitude $s_3$ and phase $\phi_3$ of the near-field signal as a function of the tip position in the $x$–$y$ plane. The numbered black dots and arcs mark the high-intensity 'hot-rings' at the nanocone sidewalls. The resonance frequencies at which the images were acquired are indicated by the labels $^U\text{TM}_{13}$, $^U\text{TM}_{14}$, and $^U\text{TM}_{15}$. All the scale bars are 300 nm.

The lack of any significant near-field contrast on the surface of the nanocones at frequencies outside of the hyperbolic Reststrahlen bands (see Supporting Information) suggests that the hot-rings were produced by HPhPs stimulated inside the nanocones. This notion gets strong support from modeling and analytical theory. In Figure 4a we present a numerically computed distribution of the tangential (to the incident) $E_t$ electric field on the surface of a nanocone of approximately the same shape and size as in Figure 3. In this simulation the system is excited solely by a plane-wave of frequency corresponding to the left column ($^UTM_{13}$) of Figure 3 incident at a 45° angle with respect to the optic axis and $p$-polarized along the $x$-axis, consistent with the s-SNOM experiment. Although no tip is included in the simulation and the relationship between the near-field signal and local field is known to be nontrivial, a good qualitative match between the calculated and the measured $s_3$ is evident. In particular, three hot-rings are clearly seen, just as in the experiment. Next, in Figure 4b we show the electric field distribution computed in a cross-section taken through the middle of the nanocone. The dominant features of this distribution are high-intensity "rays" that are emitted at the bottom edge of the nanocone and propagate along a straight line until they experience internal reflection off the sidewalls. The repeated propagation and reflection gives rise to the cross-hatch pattern of the $E$ field intensity. Comparing Figure 4a and 4b, the origin of the hot-rings becomes clear: these rings should correspond to the reflection points of the volume-confined HPhP rays. As shown in Figure 2 (b and c), a stark contrast between anomalous and specular reflection is anticipated, with only the former giving rise to the 'hot-rings' observed in the s-SNOM measurements reported here.

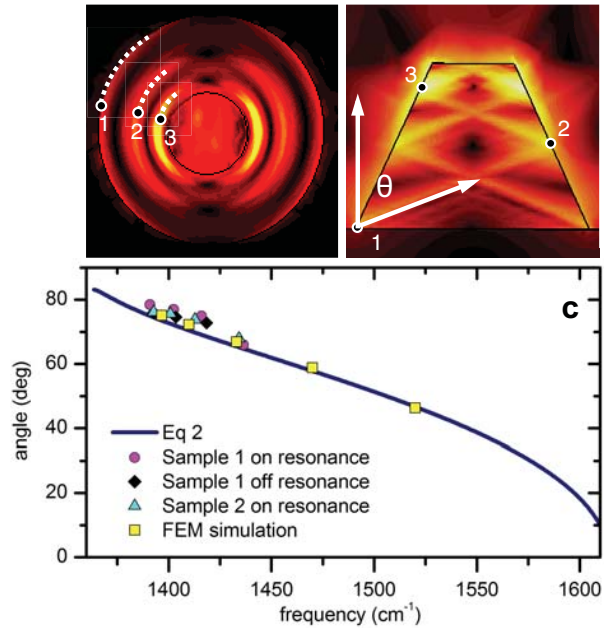

**Figure 4.** Polariton trajectories. **a** Numerically calculated distribution of tangential electric field at the surface of the nanocone (top view). The calculation was done for $^UTM_{13}$ resonant frequency. The numbered dots and arcs indicate the radial positions of the "hot" rings observed in Figure 3. **b** The same tangential electric field distribution for the $^UTM_{13}$ in the $x$–$z$ cross section of the nanocone. The hot rings arise at the sidewall reflection points of the HPhP trajectories launched at the nanocone base. **c** HPhP propagation angle $\theta$ as a function of frequency. The solid line is given by eq 2, while the symbols represent the values extracted from the near-field images (such as those in Figure 3) for the two samples, both on and off the resonance, as well as those extracted from numerical simulations.

As explained above, the polariton reflections should be anomalous, in particular the direction of incident and reflected rays should be independent of the sidewall tilt angle (see Figure 2). To test this prediction, we carried out the following data analysis. Using $\theta$ as an adjustable parameter, we simulated polariton trajectories that originate at the bottom edge of the nanocone and zigzag up to the top making angles either $\theta$ or $180° - \theta$ with respect to the z-axis. We then determined $\theta$ from the best fit among the measured radial positions of the hot-rings and thus the reflection points of such simulated trajectories (see Supporting Information) along the nanocone sidewalls. We plotted the extracted propagation angles in Figure 4c. As one can see, the data points extracted from s-SNOM plots collected from both nanocone arrays near the corresponding resonant frequencies (pink circles and blue triangles) are located very close to the theoretical curve computed from eq 2 using the dielectric permittivity model extracted from hBN flakes reported in our prior work.[3] To further verify this agreement with the predicted behavior associated with anomalous reflections, we also performed the same analysis upon the top-views of the FEM simulations at each of the resonant frequencies (yellow squares), which also agree exceptionally well with the theoretical predictions and s-SNOM experimental data. The offset between the experiment and theory, which is well within the experimental uncertainty, may be attributed to errors in determining the ring positions from the s-SNOM images and/or the inaccuracy of the measured geometrical dimensions of the nanostructures. On the other hand, the conventional model of specular reflection fails to account for the ring positions. The anomalous (Figure 2b) and specular (Figure 2c) reflection would give the same result only for cylindrical particles with vertical sidewalls or for planar slabs, as in prior s-SNOM work on hBN.[5, 14, 15, 17, 26, 27] In the former case, an elementary derivation yields that the number $N$ of hot rings would have the following dependence on the height $d_z$ and diameter $d_t$ of the cylinders:

$$N = 1 + \frac{\tan\theta}{A}, \quad A \equiv \frac{d_t}{d_z}. \quad (3)$$

This formula is also a reasonable first approximation for a pillar with tilted sidewalls if $d_t$ is understood to be its median diameter (see Supporting Information). For the sake of simplicity, we use eq 3 for the qualitative discussion below.

Importantly, the s-SNOM images taken at frequencies detuned from the resonances, but still within the U-band, also exhibit the hot-rings. These annular features can be seen in monochromatic s-SNOM plots collected at $\omega = 1418.5$ cm$^{-1}$ (Supporting Information), which is located between the $^UTM_{13}$ and $^UTM_{14}$ resonances of Sample 1. Again, the corresponding $\theta$ extracted from this and the other off-resonance s-SNOM measurements fall on the same theoretical curve, eq 2, as shown by the black diamonds in Figure 4c. Therefore, the totality of our data is consistent with the anomalous reflection/directional propagation of polariton rays with the angle $\theta(\omega)$ that varies continuously with frequency across the entire U-band. While such a directional propagation has been previously demonstrated in experiments with hBN slabs[3, 5, 13-15, 17, 26-28] the polariton reflections off slab surfaces are specular and so appear "normal." Our investigation of the sloped sidewalls gives the first evidence that this non-specular, directional reflection of HPhPs is continued regardless of the interface tilt angle.

Let us turn to the spectral response of the hBN nanoparticles. We start with the far-field reflectivity of Sample 1 (blue trace in Figure 5a). The data reveals six well-resolved peaks in the U-band and as many as three peaks in the L-band. These peaks are expected as similar resonances have been reported previously[3] and shown to be in agreement with electromagnetic simulations. The following simplified analytical theory captures the main trends exhibited by these experiments and simulations. The physical picture of the resonances are long-lived standing-wave modes of polaritons inside the nanocones. The modes are characterized by the conserved $z$-axis angular momentum $m = 0, 1, \ldots$ . In the far field, the dipole-active modes $m = 0, 1$ are observable. The $m = 0$ modes couple to the out-of-plane electric field $E_z$, while the $m = 1$ modes respond to the in-plane field $E_x$. The magnetic field of each mode is predominantly in-plane, so the modes are transverse magnetic (TM) in nature. This is the reason for the aforementioned notation $^R\text{TM}_{ml}$.[3] The strength of a particular resonance depends on its dipole matrix element, which decreases with $l$. The resonance frequencies can be found from the semi-classical quantization rule for the standing waves $k_z d_z \approx l$, $k_t d_t \approx m$, where $d_t$ is the median diameter of the particle. The approximate equality signs indicate that the expressions are valid up to $O(1)$ corrections, which depend on the precise electromagnetic boundary conditions[29] (see also Supporting Information, Sec. 4.2). Combining the semiclassical rules with the first equality in eq 2, we obtain the implicit equation for the resonant frequency $\omega_{ml}$ (eq 4 of ref 3)

$$\tan \theta(\omega_{ml}) = \frac{l + O(1)}{m + O(1)} A . \qquad (4)$$

Since $\tan \theta(\omega)$ decreases with $\omega$ in the U-band, eq 4 predicts that higher-$l$ (smaller-amplitude) resonance peaks occur at lower frequencies, whereas the trend is reversed in the L-band, in agreement with Figure 5a (the blue trace). Equation 4 can also be rewritten as the condition for the two frequency-dependent permittivities:

$$\varepsilon_t(\omega_{1l}) \approx -l^2 A^2 \varepsilon_z(\omega_{1l}) . \qquad (5)$$

With $\varepsilon_z$ being positive and nearly constant in the U-band, the right-hand side of this equation is a quadratic function of $l$. This parabolic dependence is borne out by the blue triangles in Figure 5b that represent the resonant frequencies from the same-color trace of Figure 5a replotted as $\varepsilon_t$ vs. $l$, the resonance order. Finally, substituting eq 4 into eq 3 and ignoring the $O(1)$-corrections, we find that the number of the hot rings at a given $^U\text{TM}_{1l}$ resonance should be $N = l$. The imaging data in Figure 3, which correspond to $l = 3, 4$, and $5$, are consistent with this simple rule. It should be noted that volume-confined resonances were studied theoretically[3,29] for spheroidal hBN nanoparticles; however, in that case the eigenmodes show no criss-cross patterns because spheroids lack sharp edges that launch concentrated HPhP rays.

Let us now discuss the spectral dependence of the near-field response, which is the red trace in Figure 5a. These data have been obtained from nano-FTIR measurements carried out at a fixed position of the tip above the top surface of the nanocone. They are meant to represent the overall dependence of $s_3$ on frequency, without the complications introduced by the movable hot rings. Nevertheless, the nano-FTIR spectra were fairly insensitive to the tip position along the nanocone sidewall (or top surface).

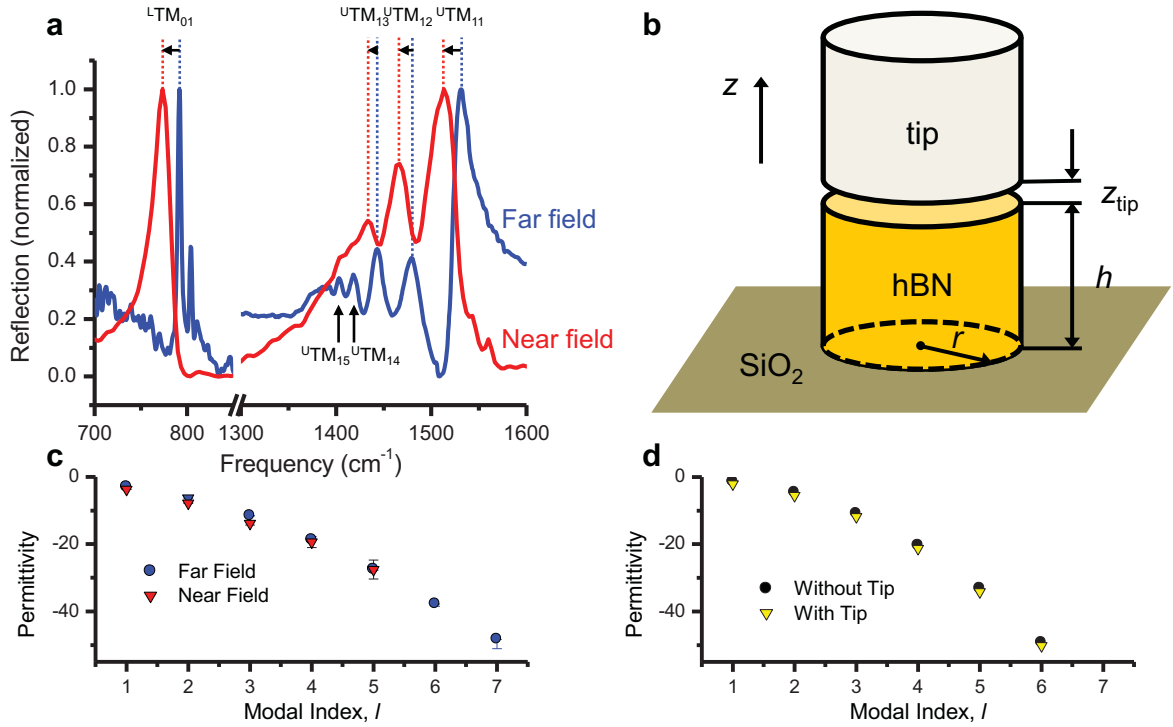

**Figure 5.** Polariton resonances. **a** The FTIR reflectance spectra of the nanocone array collected with a far-field microscope (blue) and with the nano-FTIR (red). The spectral red shifts of the near-field resonances is due to tip-sample coupling. **b** Schematics of the analytical model in which the hBN nanocone is represented by a finite-height cylinder with height-to-diameter ratio $h/2r = 12/11$, appropriate for Sample 1, and the tip is modelled as a semi-infinite metallic cylinder above it. **c** Transverse permittivity $\varepsilon_t$ vs. modal order for the far-field (blue) and near-field (red) resonances. **d** The permittivity at $^{U}TM_{1l}$ resonances computed from the model shown in b. The black symbols correspond to the case where the tip is absent; the yellow symbols are for the case where the tip is positioned $z_{tip} = 33$ nm above the sample.

Near-field observables are not subject to the dipole selection rules, so, in principle, they can reveal signatures of $m > 1$ modes[29]. Yet comparison of the red and blue traces in Figure 3a clearly indicate that no additional resonances are detected. At the same time, all the resonances observed in the near-field are red-shifted with respect to their far-field counterparts. Quantifying the magnitude of this shift illustrated that it decreases with $l$, from $\sim 15$ cm$^{-1}$ for $l = 1$ to less than 0.5 cm$^{-1}$ for $l = 6$. To analyze the origin of this red-shift we reasoned that in our sub-diffractional resonators $\omega$ may enter the resonance condition only through $\varepsilon_t$ or $\varepsilon_z$. The key quantity is the former, $\varepsilon_t$, as the latter is approximately constant across the U-band. Hence, we extracted Re($\varepsilon_t$) from the previously reported hBN dielectric function[3] at the near-field resonance conditions and plotted it (red triangles in Figure 5c) alongside their far-field counterparts (blue circles). This representation uncovers an intriguing result: the change in $\varepsilon_t$ is approximately the same for all $l$, $\varepsilon_t^{near} - \varepsilon_t^{far} \approx -1$. If we accept it as a primary fact, the smaller red-shift at larger $l$ becomes a consequence of a higher spectral dispersion of $\varepsilon_t(\omega)$ at lower $\omega$, in particular, near the bottom of the U-band where $\varepsilon_t$ becomes a very large negative value (and would diverge if hBN were lossless).

Next, it is known that resonance conditions (eq 5), of polaritonic resonators can be modified by nearby polarizable objects. It is therefore likely that the observed near-field red-shift is brought about by the presence of the metalized tip, as in s-SNOM studies of other highly resonant materials[30-33]. Unfortunately, a fully realistic modeling of this effect is challenging because of the nontrivial shapes of the hBN particle and the tip and because of widely different length scales in the problem. Therefore, to test the above hypothesis we confined ourselves to two simplified approaches. The first involved a crude analytical treatment in which the tip was approximated by a semi-infinite metallic cylinder, the hBN nanocone replaced by a finite-height cylinder (Figure 5c), and simplified boundary conditions akin to eq 4 were imposed. This calculation (see Supporting Information) correctly reproduces the sign and the order of magnitude of the shift (Figure 5d). Our second approach was the numerical solution of the Maxwell equations for the 3D system, where the hBN particle included the realistic shape, but the tip was modeled as a 500 nm diameter gold sphere. These simulations were also able to reproduce the observed red-shift with an appropriate magnitude when the sphere was placed 5 nm above the hBN nanocone. However, the magnitude of this shift was sensitive to the size, shape, and position of the 'tip' with respect to the top surface of the hBN nanocone and thus, quantitative comparison is not straightforward. Importantly, the simulated distribution of the electric near-fields inside the hBN nanoresonators in the presence of the 'tip' (gold sphere) were essentially the same as those for the far-field resonance conditions in its absence (see Supporting Information). We also attempted to experimentally perform s-SNOM measurements with a Si dielectric tip (without any metal coating), but found that the scattering signal was too weak to generate a meaningful image. In an effort to understand this further we have incorporated a dielectric tip in our analytical calculations (see Supporting Information) and observed a similar spectral shift.

These theoretical and modeling results suggest that apart from the induced spectral shift, the tip does not fundamentally alter the nature of the resonances, the anomalous reflection of the HPhPs and that the directional, volume-confined nature of the HPhP propagation are maintained. This notion is consistent with the absence of additional modes in the nano-FTIR spectra shown in Figure 5a. This thereby implies that the measured s-SNOM near-field spatial plots can be considered a good qualitative representation of the local field distribution in the absence of the tip.

With the hot-rings dominating the spatial maps of the near-field signal and the resonances being its most prominent spectral features, one may wonder if there is any deep link between the two. In fact, both theory and experiment indicate that no general relation exists. As stated, in the experiment, the rings persist even at frequencies away from the resonance condition. On the theory side, where one is free to consider particles of arbitrary shape, one generically finds resonances, but no rings.[29] What may be confusing is the existence of the simple relation between the number of rings $N$ on resonance and the resonance order $l$. However, from the derivation it is clear that this relation may not be exact even for cylinders.

Our work provides the first experimental evidence that the predicted permittivity-dependent propagation angle of HPhPs is maintained even after reflection from non-specular surfaces. This was determined from s-SNOM spatial mapping of hBN nanocones with sloped sidewalls that exhibited a series of "hot-rings" of increased near-field amplitude and nodes in the phase. Through comparison with electromagnetic simulations of these structures, it was determined that these rings resulted from the anomalous reflections of the volume-confined HPhPs within the nanocones. By comparing the extracted propagation angles (with respect to the optical axis) with those predicted for HPhPs, it was verified that the highly directional propagation is maintained regardless of the volumetric shape. Thus, as the hyperbolic nature of hBN may enable simplified hyperlens designs,[4, 5] this work ensures that the predicted propagation of HPhPs inside the material is robust against variations of the sample geometry or imperfect surface structure. Furthermore, hyperbolic materials have been highlighted as being potentially useful for a wide array of other applications ranging from quantum nanophotonics,[11] to narrow-band infrared emitters, detectors,[5] and thermal sources.[34-36] The more complete understanding of how the HPhP modes behave and how they are influenced by their environment is pertinent to future design and incorporation of HMs in nanophotonic applications. Another promising future direction is integrating hBN with other atomically thin materials into van der Waals heterostructures[37] and metamaterials. It has recently been demonstrated that through the formation of electromagnetic hybrids[38, 39] with plasmonic materials such as graphene, the hyperbolic dispersion of hBN can be actively controlled.[15, 26] The results reported here suggest that such electric-field control of the hyperbolic dispersion will also result in a direct control of the HPhP propagation angle,[40] which could be employed in the design of nanophotonic logic or sub-diffractional optics.[5]

■ ASSOCIATED CONTENT

**Supporting Information**

A detailed description of experimental procedures, characterization data, data fitting procedures, and theoretical modeling. This material is available free of charge via the Internet at http://pubs.acs.org/.

■ AUTHOR INFORMATION

**Corresponding Author**

*E-mail: joshua.caldwell@nrl.navy.mil

■ ACKNOWLEDGEMENT

A.J.G. and C.T.E. acknowledge support from the National Research Council (NRC) NRL Postdoctoral Fellowship Program. Funding for J.D.C., O.J.G. and J.G.T. was provided via the Office of Naval Research and distributed by the Nanoscience Institute at the Naval Research Laboratory. A.V.K. and K.S.N. acknowledge support from the EPSRC (UK), the Royal Society (UK), European Research Council, and EC-FET European Graphene Flagship. Research at UCSD is supported by the grants OBR- N00014- 15-1- 2671. Development of nano-infrared instrumentation is supported by ARO w911NF-13- 1-0210 and AFOSR FA9550-15- 1-0478. D.N.B is the Moore Investigator in Quantum Materials EPIQS program GBMF4533.

■ METHODS

**Sample Preparation**. Crystals of hBN were grown using a high-pressure/high temperature method[41, 42] and were exfoliated and placed onto a high resistivity silicon substrate using standard exfoliation techniques. Atomic-force microscopy was used to determine flake thickness and select suitably sized flakes for patterning. After indexing flake position, the sample was coated with a bilayer of PMMA and patterned in a standard electron beam lithography process, followed by an electron beam deposition of an Al hard mask followed by a standard lift off procedure. The hBN nanostructures were fabricated via reactive ion etching in an oxygen environment. The remaining metal was then removed with wet chemical etchants. Further details of the fabrication process can be found in ref 4.

**Far-field FTIR Spectroscopy**. FTIR reflection spectra were collected using a Bruker Vertex 80V FTIR spectrometer with Hyperion 1000 microscope attachment. The incident light was directed to the sample and the reflected light collected through an all-reflective 15×, 0.452 NA, reverse Cassegrain objective. A 50 $\mu$m aperture, located at the image plane of the microscope, is used to limit the collection area to only regions with or without the hBN nanoresonators. The collected light was detected using a liquid nitrogen cooled mercury cadmium telluride (MCT) detector. In all cases, the spectra were collected in reference to a gold mirror to enable quantitative measurements.

■ REFERENCES

1. Poddubny, A.; Iorsh, I.; Belov, P.; Kivshar, Y. *Nature Photonics* **2013,** 7, 948-957.
2. Yang, X.; Yao, J.; Rho, J.; Xiaobo, Y.; Zhang, X. *Nature Photonics* **2012,** 6, 450-453.
3. Caldwell, J. D.; Kretinin, A.; Chen, Y.; Giannini, V.; Fogler, M. M.; Francescato, Y.; Ellis, C.; Tischler, J. G.; Woods, C.; Giles, A. J.; Hong, M.; Watanabe, K.; Taniguchi, T.; Maier, S. A.; Novoselov, K. S. *Nature Communications* **2014,** 5, 5221.
4. Dai, S.; Ma, Q.; Andersen, T.; McLeod, A. S.; Fei, Z.; Liu, M. K.; Wagner, M.; Watanabe, K.; Taniguchi, T.; Thiemens, M.; Keilmann, F.; Jarillo-Herrero, P.; Fogler, M. M.; Basov, D. N. *Nature Communications* **2015,** 6, 6963.
5. Li, P.; Lewin, M.; Kretinin, A. V.; Caldwell, J. D.; Novoselov, K. S.; Taniguchi, T.; Watanabe, K.; Gaussman, F.; Taubner, T. *Nature Communications* **2015,** 6, 7507.
6. Yang, K. Y.; Giannini, V.; Bak, A. O.; Amrania, H.; Maier, S. A.; Phillips, C. C. *Physical Review B* **2012,** 86, 075309.
7. Liu, Z.; Lee, H.; Xiong, Y.; Sun, C.; Zhang, X. *Science* **2007,** 315, 1686.
8. Xiong, Y.; Liu, Z.; Zhang, X. *Applied Physics Letters* **2009,** 94, (20), 203108.
9. Noginov, M.; Lapine, M.; Podolskiy, V. A.; Kivshar, Y. *Opt. Express* **2013,** 21, (12), 14895-14897.
10. Biehs, S.-A.; Tschikin, M.; Ben-Abdallah, P. *Physical Review Letters* **2012,** 109, 104301.
11. Cortes, C. L.; Newman, W.; Molesky, S.; Jacob, Z. *Journal of Optics* **2012,** 14, 063001.
12. Guo, Y.; Newman, W.; Cortes, C. L.; Jacob, Z. *Advances in OptoElectronics* **2012,** 2012, 452502.
13. Ishii, S.; Kildishev, A. V.; Narimanov, E. E.; Shalaev, V. M.; Drachev, V. P. *Laser and Photonics Reviews* **2013,** 7, 265-271.
14. Dai, S.; Fei, Z.; Ma, Q.; Rodin, A. S.; Wagner, M.; McLeod, A. S.; Liu, M. K.; Gannett, W.; Regan, W.; Thiemens, M.; Dominguez, G.; Castro Neto, A. H.; Zettl, A.; Keilmann, F.; Jarillo-Herrero, P.; Fogler, M. M.; Basov, D. N. *Science* **2014,** 343, (6175), 1125-1129.
15. Dai, S.; Ma, Q.; Liu, M. K.; Anderson, T.; Fei, Z.; Goldflam, M.; Wagner, M.; Watanabe, K.; Taniguchi, T.; Thiemens, M.; Keilmann, F.; Jannsen, G.; Zhu, S. E.; Jarillo-Herrero, P.; Fogler, M. M.; Basov, D. N. *Nature Nanotechnology* **2015,** 10, (8), 682-686.
16. Jia, Y.; Zhao, H.; Guo, Q.; Wang, X.; Wang, H.; Xia, F. *ACS Photonics* **2015,** 2, (7), 907-912.

17. Shi, Z.; Bechtel, H. A.; Berweger, S.; Sun, Y.; Zeng, B.; Chenhao, J.; Chang, H.; Martin, M. C.; Raschke, M. B.; Wang, F. *ACS Photonics* **2015,** 2, (7), 790-796.
18. Xu, X. G.; Ghamsari, B. G.; Jiang, J.-H.; Gilburd, L.; Andreev, G. O.; Zhi, C.; Bando, Y.; Golberg, D.; Berini, P.; Walker, G. C. *Nature Communications* **2014,** 5, 4782.
19. Xu, X. G.; Tanur, A. E.; Walker, G. C. *Journal of Physical Chemistry A* **2013,** 117, 3348-3354.
20. Caldwell, J. D.; Lindsey, L.; Giannini, V.; Vurgaftman, I.; Reinecke, T.; Maier, S. A.; Glembocki, O. J. *Nanophotonics* **2015,** 4, (1), 44-68.
21. Maier, S. A., *Plasmonics: Fundamentals and Applications*. Berlin, 2007.
22. Feng, K.; Steyer, W.; Islam, S. M.; Verma, J.; Jena, D.; Wasserman, D.; Hoffman, A. J. *Applied Physics Letters* **2015,** 107, 081108.
23. Caldwell, J. D.; Glembocki, O. J.; Sharac, N.; Long, J. P.; Owrutsky, J. O.; Vurgaftman, I.; Tischler, J. G.; Bezares, F. J.; Wheeler, V.; Bassim, N. D.; Shirey, L.; Francescato, Y.; Giannini, V.; Maier, S. A. *Nano Letters* **2013,** 13, (8), 3690-3697.
24. Chen, Y.; Francescato, Y.; Caldwell, J. D.; Giannini, V.; Maß, T. W. W.; Glembocki, O. J.; Bezares, F. J.; Taubner, T.; Kasica, R.; Hong, M.; Maier, S. A. *ACS Photonics* **2014,** 1, (8), 718-724.
25. Wang, T.; Li, P.; Hauer, B.; Chigrin, D. N.; Taubner, T. *Nano Letters* **2013,** 13, (11), 5051-5055.
26. Woessner, A.; Lundeberg, M. B.; Gao, Y.; Principi, A.; Alonso-Gonzalez, P.; Carrega, M.; Watanabe, K.; Taniguchi, T.; Vignale, G.; Polini, M.; Hone, J.; Hillenbrand, R.; Koppens, F. H. L. *Nature Materials* **2014,** 14, 421-425.
27. Yoxall, E.; Schnell, M.; Nikitin, A. Y.; Txoperena, O.; Woessner, A.; Lundeberg, M. B.; Casanova, F.; Hueso, L. E.; Koppens, F. H. L.; Hillenbrand, R. *Nature Photonics* **2015,** 9, (674-678).
28. Fisher, R. K.; Gould, R. W. *Physical Review Letters* **1969,** 22, 1093-1095.
29. Sun, Z.; Gutiérrez-Rubio, Á.; Basov, D. N.; Fogler, M. M. *Nano Letters* **2015,** 15, (7), 4455-4460.
30. Jiang, B.-Y.; Zhang, L. M.; Castro Neto, A. H.; Basov, D. N.; Fogler, M. M. *Journal of Applied Physics* **2016,** 119, (5), 054305.
31. Amarie, S.; Keilmann, F. *Physical Review B* **2011,** 83, (4), 045404.
32. McLeod, A. S.; Kelly, P.; Goldflam, M. D.; Gainsforth, Z.; Westphal, A. J.; Dominguez, G.; Thiemens, M. H.; Fogler, M. M.; Basov, D. N. *Physical Review B* **2014,** 90, (8), 085136.
33. Zhang, L. M.; Andreev, G. O.; Fei, Z.; McLeod, A. S.; Dominguez, G.; Thiemens, M.; Castro-Neto, A. H.; Basov, D. N.; Fogler, M. M. *Physical Review B* **2012,** 85, (7), 075419.
34. Schuller, J. A.; Taubner, T.; Brongersma, M. L. *Nature Photonics* **2009,** 3, 658-661.
35. Greffet, J.-J.; Carminati, R.; Joulain, K.; Mulet, J. P.; Mainguy, S. P.; Chen, Y. *Nature* **2002,** 416, (6876), 61-64.
36. Jacob, Z. *Nature Materials* **2014,** 13, 1081-1083.
37. Geim, A. K.; Grigorieva, I. V. *Nature* **2013,** 499, 419-425.
38. Caldwell, J. D.; Vurgaftman, I.; Tischler, J. G.; Glembocki, O. J.; Owrutsky, J. C.; Reinecke, T. L. *Nature Nanotechnology* **2016,** 11, 9-15.
39. Caldwell, J. D.; Novoselov, K. S. *Nature Materials* **2015,** 14, 364.
40. Kumar, A.; Low, T.; Fung, K. H.; Avouris, P.; Fang, N. X. *Nano Letters* **2015,** 15, (5), 3172-3180.
41. Watanabe, K.; Taniguchi, T.; Kanda, H. *Nature Materials* **2004,** 3, 404-409.
42. Taniguchi, T.; Watanabe, K. *Journal of Crystal Growth* **2007,** 303, (2), 525-529.

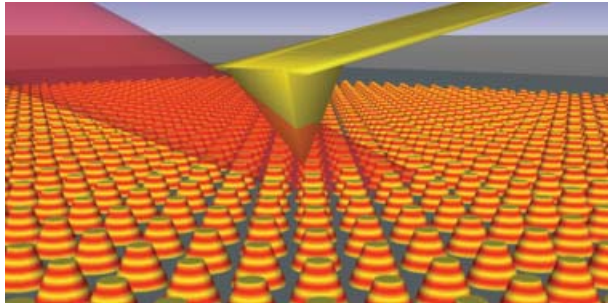
FOR TABLE OF CONTENTS ONLY